\documentclass[reprint,superscriptaddress,amsmath,amssymb,aps,prd,twocolumn,floatfix,nofootinbib]{revtex4-2}

\usepackage{hyperref,url}
\usepackage[hyphenbreaks]{breakurl}
\usepackage{graphicx,graphics,amsfonts,amsmath,mathrsfs,amssymb,bm,psfrag,natbib,dcolumn,textcase,enumerate,sidecap,listings,tabularx,mathtools,multirow}
\usepackage{float,color}
\usepackage{xcolor}
\usepackage[normalem]{ulem}

\usepackage{scrextend}

\allowdisplaybreaks

\newcommand{\be}{\begin{equation}}
\newcommand{\ee}{\end{equation}}

\newcommand{\bs}{\begin{subequations}}
\newcommand{\es}{\end{subequations}}

\begin{document}
\title{Testing general relativity via direct measurement of black hole kicks}
\author{Parthapratim Mahapatra}\email{ppmp75@cmi.ac.in}
\affiliation{Chennai Mathematical Institute, Siruseri, 603103, India}
\author{Marc Favata}
\affiliation{Department of Physics \& Astronomy, Montclair State University, 1 Normal Avenue, Montclair, New Jersey 07043, USA}
\author{K. G. Arun}
\affiliation{Chennai Mathematical Institute, Siruseri, 603103, India}
\affiliation{Institute for Gravitation and the Cosmos, Penn State University, University Park, Pennsylvania 16802, USA}
\date{\today}
\begin{abstract}
Asymmetric emission of gravitational waves during a compact binary coalescence results in the loss of linear momentum and a corresponding ``kick'' or recoil on the binary's center of mass. This leads to a direction-dependent Doppler shift of the ringdown gravitational waveform. We quantify the measurability of the kick imparted to the remnant black hole in a binary black hole merger. Future ground- and space-based gravitational-wave detectors will measure this effect to within $\sim 2\%$ to $\sim 30\%$ for a subset of their expected observed sources. Certain binary configurations in the LISA band may allow a sub-percent-level measurement of this effect.
This \emph{direct} measurement of black hole kicks can also facilitate a novel test of general relativity based on linear momentum balance. We formulate this \emph{kick consistency} test via measurement of a null variable that quantifies the difference between the inferred kick (using numerical relativity) and that observed via the Doppler-shifted ringdown signal. This null variable can be constrained (at 90\% confidence) to $\sim 10\%$ to $30\%$ with Cosmic Explorer and to $\sim 3\%$ to $12\%$ with LISA. 
\end{abstract}
\maketitle
\setlength\abovedisplayskip{0pt}
\setlength\belowdisplayskip{0pt}
\section{Introduction}
The emission of gravitational waves (GWs) from a compact binary leads to the loss of energy, angular momentum, and linear momentum from the system. Linear momentum flux arises from the asymmetric emission of GWs in binaries with unequal component masses or spin angular momenta \cite{BoR61, Peres62, Bek73, Fitchett83, Wi92, K95, Favata:2004wz, BQW05}. This leads to the recoil or ``kick'' of the binary's center of mass following coalescence. The final kick following the merger of two black holes (BHs) can range from $\sim 10\,{\rm km/s}$ to $\sim 5000 \,{\rm km/s}$ \cite{Gonzalez:2006,Campanelli:2007ew,Campanelli:2007cga,Gonzalez:2007hi,Lousto:2007,Tichy:2007hk, Lousto:2011kp, Lousto:2012su, Lousto:2013,Lousto:2019lyf}, with the magnitude and direction depending strongly on the binary mass ratio and the component spin magnitudes and directions. If these kicks exceed the escape speeds of galaxies (for supermassive BHs) or star clusters (for stellar mass BHs), the BH merger remnant will be ejected \cite{Merritt04}, prohibiting the pairing with another BH and a subsequent merger~\cite{Gerosa:2017kvu,Fishbach:2017dwv,Fragione:2017blf,Antonini:2018auk,Antonini:2016gqe,Mahapatra:2022ngs,Mahapatra:2021hme}. This impacts our understanding of BH growth~\cite{Miller:2001ez,Merritt04, Bellovary:2019nib}.

The purposes of this paper are to (i) quantify the precision that BH kicks can be \emph{directly measured} from the GW signal (as opposed to inferring the kick via propagated constraints on binary parameters) and (ii) provide a proof-of-principle demonstration of a new test of general relativity (GR) based on this direct measurement.

Given the intrinsic parameters of a binary BH (BBH) system (the mass ratio $q\equiv m_1/m_2\geq 1$ and dimensionless spin angular momenta $\vec{\chi}_{1,2}$), numerical relativity (NR) simulations predict the final mass $M_f$, dimensionless spin $\vec{\chi}_f$, and kick velocity $\vec{V}_{\rm kick}$ of the BH merger remnant. (Here $m_1$ and $m_2$ are the primary and secondary source-frame masses of the binary components; we assume circular orbits and $G=c=1$.) Fitting formulas to NR simulations or surrogate models map the inferred binary component parameters to these remnant properties [see, e.g., Refs.~\cite{Campanelli:2004,Campanelli:2007ew,Campanelli:2007cga,Lousto:2011kp, Lousto:2012su, Lousto:2013,Lousto:2019lyf,Gerosa:2018qay,Varma:2018aht} for the mapping $\vec{V}_{\rm kick}(q,\vec{\chi}_{1},\vec{\chi}_{2}$)]. Hence, estimating binary parameters during the inspiral portion of the GW signal provides a prediction for the remnant mass, spin, and kick. 

\begin{figure}[t!]
    \centering
    \includegraphics[scale=0.3]{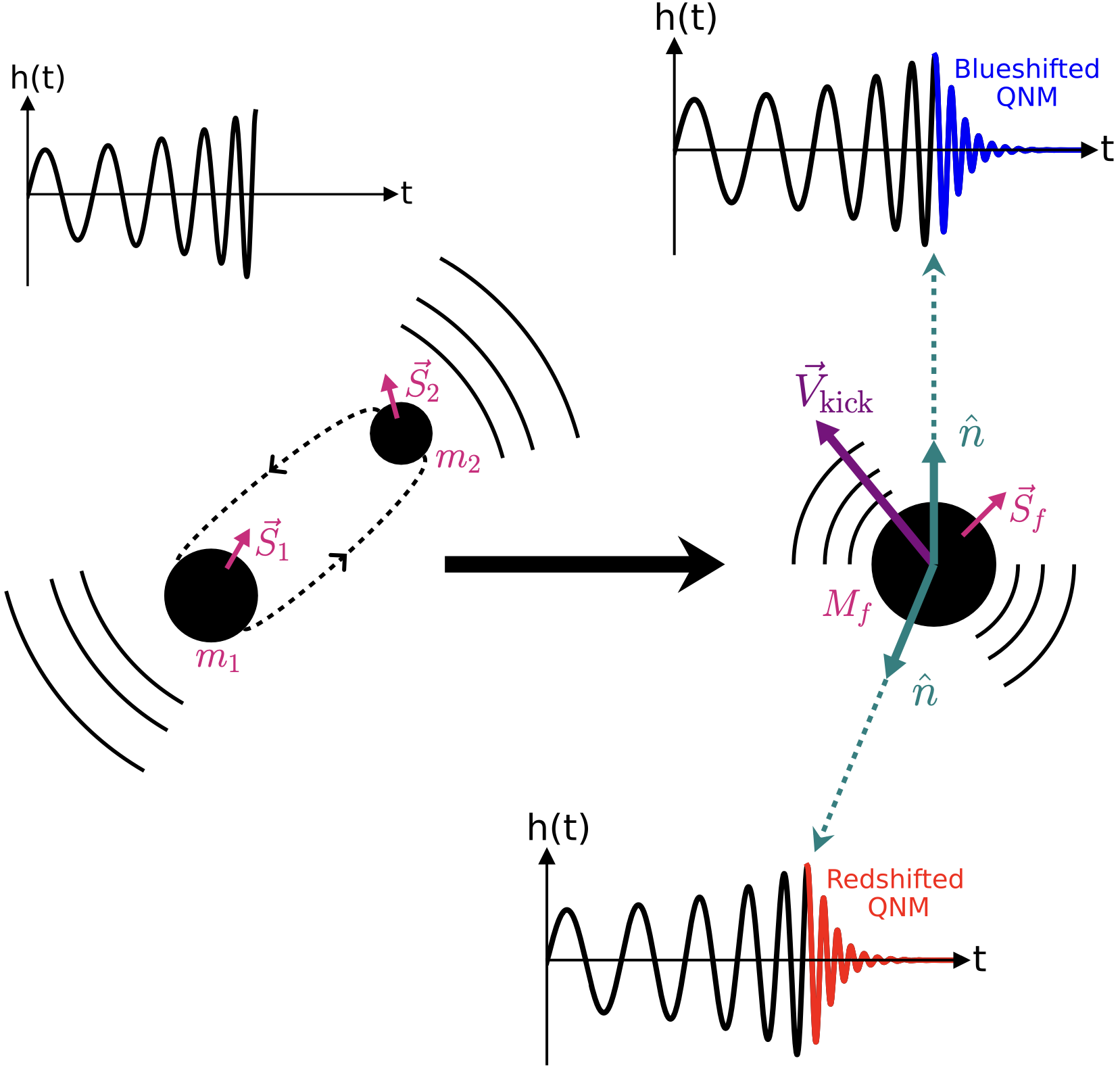}
    \caption{\label{fig:kickcartoon}{Illustration of kick's imprint on the gravitational-wave signal from a black hole binary with component masses $m_1$, $m_2$ and spins $\vec{S}_1$, $\vec{S}_2$. During the inspiral (left) the slowly growing Doppler shift due to radiated linear momentum is negligible. A burst of energy, angular momentum, and linear momentum during the merger and ringdown phases (right), yields a remnant black hole with mass $M_f$, spin $\vec{S}_f$, and recoil velocity $\vec{V}_{\rm kick}$. The kick imprints a Doppler shift on the merger/ringdown waveform which depends on the direction $\hat{n}$ to the observer via $V_r\equiv \hat{n} \cdot \vec{V}_{\rm kick}$.}}
\end{figure}

\begin{figure}[t]
    \centering
    \includegraphics[scale=0.46]{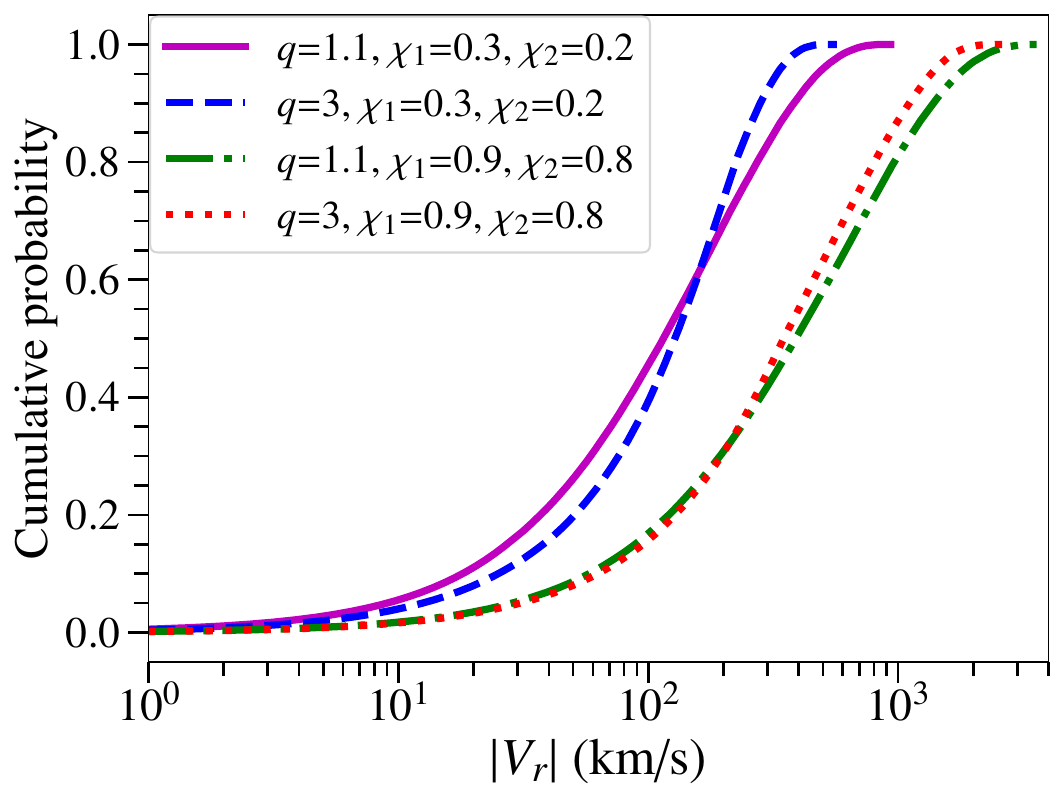}
    \caption{\label{fig:LOSkickCDF}Cumulative probability distributions for the LOS kick for four binary black hole configurations (see text). Curves terminate at the highest sampled LOS kick.}
\end{figure}

Separately, it is possible to directly measure the remnant mass, spin, and kick from the final stages of a GW signal. It has long been understood that the quasinormal mode (QNM) frequencies and damping times encode the remnant's mass $M_f$ and spin magnitude $|\vec{\chi}_f|$. That the kick could also be extracted from the QNM signal was first pointed out by Favata \cite{Favata:2008ti}.\footnote{The kick also contributes a memory effect to the GW signal that scales like $h_{\rm mem., kick} \sim M_f V_{\rm kick}^2/D_L$, where $D_L$ is the source luminosity distance \cite{Favata:2008ti}. There is additionally an aberration effect \cite{Torres-Orjuela:2020cly} that is smaller than the Doppler shift but also of order $V_{\rm kick}$. Both of these effects are ignored here.} For a widely separated binary at rest relative to the observer, the small radiated linear momentum during the inspiral imparts only a small-amplitude oscillation to the binary's center of mass \cite{Favata:2004wz}. But the larger linear momentum loss during the merger/ringdown phase of coalescence results in a nearly step-function kick imparted rapidly during the merger. Hence, the inspiral signal has a negligible Doppler shift, while the ringdown frequencies $f_{\rm RD}$ are shifted via
\be
\label{eq:doppler}
f_{\rm RD, Dopp.}= \frac{f_{\rm RD}}{\gamma (1 - \hat{ n} \cdot \vec{V}_{\rm kick})} \approx f_{\rm RD} \left(1+V_r\right),
\ee
where $\gamma$ is the Lorentz factor, ${\hat  n}$ is the unit vector along the direction of propagation from the source to the observer, and $V_r \equiv \hat{ n} \cdot \vec{V}_{\rm kick}$ is the line-of-sight (LOS) component of the recoil (with kicks away from the observer producing a redshift).\footnote{The cosmological redshift is a separate effect imparting a constant frequency shift to all phases of the GW signal. For binaries within a triple system, there arises an oscillatory Doppler shift in the GW signal \cite{Yunes:2010sm, Meiron:2016ipr, Inayoshi:2017hgw, Chamberlain:2018snj, Randall:2018lnh, Robson2018, Wong:2019hsq, Torres-Orjuela:2020cly,Vijaykumar:2023tjg} modulated on the timescale associated with the third body's orbital period. Both effects are distinct from the step-function-like shift imparted by the GW recoil.} (See Fig.\ref{fig:kickcartoon} for an illustration.)  Since BH kicks are nonrelativistic ($V_r<0.01, \gamma \approx 1$, c.f.~Fig.~\ref{fig:LOSkickCDF}), we apply the approximation on the right-hand side of Eq.~\eqref{eq:doppler}. The QNM damping times $\tau_{\rm RD}$ are similarly Doppler-shifted via $\tau_{\rm RD, Dopp.}^{-1} \approx \tau_{\rm RD}^{-1} (1+V_r)$.

Favata~\cite{Favata:2008ti} crudely estimated that the LOS kick component could be measured to a fractional accuracy $\sigma_{V_r}/V_{r} \sim 2/(\rho V_{r})$, where $\rho$ is the signal-to-noise ratio (SNR). A more detailed analysis was performed by Gerosa and Moore \cite{Gerosa:2016vip} via the \emph{mismatch} between waveforms with and without a kick-induced Doppler shift. They found that supermassive BH (SMBH) kicks $\gtrsim 500\, {\rm km/s}$ could be detected by LISA. 

Other works considered inference of the kick via propagation of a binary's estimated parameters through NR fitting formulas or surrogate waveform models~\cite{Gerosa:2018qay,Varma:2018aht,Varma:2019csw}. For example, Ref.~\cite{Varma:2020nbm} demonstrated the prospects of inferring the kicks of GW150914 and GW170729, while Ref.~\cite{GW190521ApJL} used this approach to constrain the kick of GW190521. Mahapatra {\it et}~{\it al}.~\cite{Mahapatra:2021hme} investigated the inferred kicks for all BBHs in GWTC-2~\cite{GWTC2}, finding that GW190814 has the best constrained kick ($74^{+10}_{-7} {\rm km/s}$), while Ref.~\cite{Varma:2022pld} showed that the precessing system GW200129~\cite{Hannam:2021pit} has a large inferred kick of $1542^{+747}_{-1098}\, {\rm km/s}$. Higher-order waveform modes were also used to infer the kick direction \cite{CalderonBustillo:2022ldv} and LOS component~\cite{CalderonBustillo:2018zuq}. In particular, we note that Ref.~\cite{CalderonBustillo:2018zuq} has shown that the inclusion of higher modes allows for a more precise inference of the LOS kick than an analysis that relies solely on a measurement of the Doppler-shifted ringdown modes as proposed in \cite{Favata:2008ti} and explored in \cite{Gerosa:2016vip}. Moreover, Ref.~\cite{Varma:2020nbm} argued that the LOS kick-induced Doppler effect should be accounted for in the waveform to avoid systematic biases in ringdown tests of GR.

In contrast to the above works on inferring the kick, here we further the analyses of Refs.~\cite{Favata:2008ti,Gerosa:2016vip} by focusing on assessing the accuracy with which the LOS kick component ($V_r$) can be \emph{directly measured} (rather than inferred) via third generation (3G) ground-based detectors~\cite{CEScience,CEScience19,ETScience10} and LISA~\cite{LISAScience17}. This measured Doppler shift serves as an \emph{effective parameter} that directly probes the velocity imparted to the remnant at the merger time, regardless of the mechanism or theory of gravity. (For example, a fortuitous physical collision with a third object just at the merger would also impart a Doppler shift.)

We also propose that this direct measurement of the LOS kick can be used to formulate a new \emph{kick consistency test} of GR. In this test the inspiral waveform (possibly combined with information from the merger/ringdown)  constrains the binary component parameters (masses and spins) and the source-to-observer direction ($\hat{n}$). Propagating these constraints through NR fits leads to posterior probability constraints on $\vec{V}_{\rm kick}$ and the LOS component $V_r$. This \emph{prediction} of $V_r$ is compared with the measurement of $V_r$ from the ringdown Doppler shift, which provides a separate constraint. Because a Doppler shift imparted to the signal is a generic physical process associated with a sudden change in the center-of-mass velocity, consistency between these two constraints would provide the first direct evidence that GWs carry linear momentum as GR predicts (i.e., the observed Doppler shift is consistent with kicks predicted by GR while applying linear momentum conservation). From a fundamental physics standpoint, verifying the global conservation of linear momentum is as important as (and complementary to) verifying energy and angular momentum conservation (as in current inspiral/merger/ringdown tests~\cite{Ghosh2016,Hughes:2004vw}). An inconsistency between predicted and measured values of $V_r$ could indicate a GR violation.  

In the remainder of this paper, we first investigate the range of plausible LOS kick velocities, accounting for spin precession. We then modify a phenomenological frequency-domain (nonprecessing) waveform model ({\tt IMRPhenomD}~\cite{Husa:2015iqa, Khan:2015jqa}) to include the kick-induced Doppler shift and determine how well $V_r$ can be constrained via a Fisher matrix analysis.  
Lastly, we perform a separate parameter estimation calculation using a precessing waveform model ({\tt IMRPhenomXPHM}) \cite{IMRPhenomXPHM} to determine the system parameters, direction to the observer $\hat{n}$, and the \emph{inferred} kick vector $\vec{V}_{\rm kick}$. These analyses are combined to determine the constraints on a null variable $\delta \hat{V}_r$ that quantifies the consistency between the two determinations of the LOS kick component $V_r$. 

\section{Expected line-of-sight kicks}
While BH kicks as large as $\sim 5000 \,{\rm km/s}$ are possible~\cite{Campanelli:2007cga,Lousto:2011kp,Lousto:2019lyf}, these large kicks rely on optimal spin configurations. Figure~\ref{fig:LOSkickCDF} explores the expected distribution of LOS kick components $V_{r}$ via NR surrogate models \cite{Varma:2018aht,Varma:2019csw}. We consider four BBH systems with two mass ratios ($q=1.1$ or $3$) and two sets of spin magnitudes: high spins ($\chi_1=0.9,\, \chi_2=0.8$) and low spins ($\chi_1=0.3, \, \chi_2=0.2$). The spin orientations and detector position angles ($\hat{n}$) are isotropically sampled $\sim 10^5$ times over the sphere, with Fig.~\ref{fig:LOSkickCDF} showing the resulting cumulative probability distribution. High-spin binaries (and to a lesser extent, those with near-equal masses) preferentially produce larger LOS kicks. For example, only $9\%$ of the low-spin, $q=1.1$ binaries yield LOS kicks exceeding $400\, {\rm km/s}$; while $13\%$ of high-spin, $q=3$ binaries achieve $V_r\geq 1000$ km/s.

\section{Doppler-shifted waveform and parameter estimation}
As with an electromagnetic source, a GW source moving relative to an observer will display a Doppler-shifted frequency spectrum as compared to the same source at rest. As a BBH inspirals and merges, anisotropies in the emitted GWs carry away linear momentum, resulting in a recoil of the binary or remnant that quickly builds and saturates during the merger/ringdown. The resulting ringdown signal will be composed of a sum of damped sinusoids with frequencies and damping times that depend on $M_f$, $\chi_f$, and $V_r$. Hence, the spectral amplitude of the signal will depend on the direction of the observer relative to the direction of $\vec{V}_{\rm kick}$ via the $V_r$ dependence. In other words, if kicks did not occur, the QNM spectrum would be different, and this spectrum depends on the observer's location.

The decomposition of the observed signal into a mode sum depends on the chosen basis. It is a standard practice in NR simulations to decompose the waveform plus and cross polarizations $h_{+,\times}$ onto the basis of spin-weighted spherical harmonic functions via
\be
h_{+}- i h_{\times} = \sum_{l\geq 2, |m|\leq \ell} h_{\ell m}(T) {}_{-2}Y_{\ell m}(\Theta,\Phi)\,.
\ee
Here the angles $(\Theta,\Phi)$ are equivalent to the unit vector $\hat{n}$ pointing to the observer from the source frame. We note that the modes $h_{\ell m}(T)$ depend only on the retarded time $T$ measured at the observer and not on any angles. Hence, any kick-related angle dependence of the GW signal's frequency spectrum (i.e., the Doppler shift) must be encoded in the relative amplitudes of the $h_{\ell m}$ modes. The relationships to the QNMs of a Kerr BH with mass $M_f$ and spin $\chi_f$ are further complicated by the fact that those QNMs are defined relative to the basis of spin-weighted \emph{spheroidal} harmonics. Hence, the standard QNMs in the spheroidal harmonic basis become coupled when a spherical harmonic basis is used (see, e.g., \cite{MitmanPhysRevD.106.084029} and references therein).

Considering the above, we defer a precise analysis of how the kick and Doppler shift affect each mode to future work. Here, we wish to perform a ``first-cut'' analysis that estimates the measurability of the kick via the Doppler-shifted waveform. Further, we also consider the case where the kick may differ from the predictions of GR, either because (i) GR is not the correct theory, or (ii) (as a point of principle rather than practice) the kick is caused by an astrophysical mechanism (i.e., an interaction or collision with a third body). To do this, we perform a ``by-hand'' modification of the GW waveform (discussed below), considering a model dominated only by the $(\ell, m)=(2,\pm2)$ modes. While the GW-induced kick in GR is not a free parameter but a prediction that follows from the initial system parameters, here we treat $V_r$ as an \emph{effective} parameter that sets the strength of the kick's influence on the post-merger signal. Measurement of $V_r$ allows us to check agreement with the predicted GR value (see Sec.~\ref{sec:GRtest} below).

To incorporate the kick, we modify the nonprecessing phenomenological {\tt IMRPhenomD}~\cite{Husa:2015iqa, Khan:2015jqa} waveform. That frequency-domain model splits the GW signal into three phases (inspiral, intermediate, and merger/ringdown) that are ``stitched'' together via step functions. Since the kick primarily accumulates during the final merger/ringdown, we only modify that phase by replacing the real and imaginary components of the ringdown frequency: $(f_{\rm RD}, f_{\rm damp}) \rightarrow (f_{\rm RD}, f_{\rm damp}) (1+V_r)$. Since this model contains only a single mode, a kick-induced Doppler shift cannot be present in the mode (prior to modification), and the issues of mode-mixing and understanding how the kick affects each mode become moot. We are working in an approximation whereby the entirety of the kick-induced Doppler shift is encapsulated in a single effective parameter $V_r$ that shifts the frequencies and damping times of the dominant (and in this case, only) mode of the signal. In other words, in the context of a waveform model containing only the $h_{2,\pm 2}$ modes, the angle-dependent kick-induced Doppler shift can only be incorporated via the ansatz above. Future work applied to multimode waveform models can relax this assumption.

We modify the {\tt IMRPhenomD} model (as opposed to one incorporating precession, such as {\tt IMRPhenomXPHM} \cite{IMRPhenomXPHM} used below) for several additional reasons. First, in this section we are interested in treating $V_r$ as a free parameter 
whose measurement precision we wish to determine; hence, the dependence of $V_r$ on the binary parameters is not needed for this part of the analysis. Further, while 3G detectors are much more sensitive to the effects of precession \cite{vitale2016PRD}, the precessional dynamics primarily affects the inspiral waveform (where the kick is negligible) and has minimal effect on the dominant ringdown mode that most contributes to the post-merger SNR~\cite{Meidam2014,Kamaretsos2012}. While fully-precessing waveforms have a more complex mode structure, the $(2,2)$ mode still dominates, especially for the near-equal-mass binaries that we consider below. Inclusion of higher modes would likely lead to tighter constraints on $V_r$ than our conservative estimates. Also, Ref.~\cite{Gerosa:2016vip} showed that the kick's detectability is insensitive to the waveform model (precessing vs.~nonprecessing) and depends primarily on the ringdown SNR. Lastly, the {\tt IMRPhenomD} waveform is easily modified. This suffices for our proof-of-principle study. Note that our formulas for $\vec{V}_{\rm kick}$ and $V_r$ fully account for precession; and our inputs to {\tt IMRPhenomD} consistently include only the $z$-components of our precessing spin configurations. (The \emph{predicted} LOS component is computed below using a fully-precessing waveform.)  Future work could use models (not yet available) that consistently include both precession and the Doppler shift in all analysis stages.

To determine the measurement precision of $V_r$ we apply a Fisher matrix analysis \cite{Cramer46,Rao45,CF94} to the modified {\tt IMRPhenomD} waveform. Given a detector's noise power spectral density $S_n(f)$ (PSD), the Fisher matrix provides a simple approach to assess lower bounds on parameter measurement precision in the case of stationary, Gaussian noise and the high-SNR limit. For a frequency-domain waveform model $h(f; \lambda^a)$ that depends on a set of parameters $\lambda^a$, the Fisher matrix is 
\be
\label{eq:fisher}
\Gamma_{ab}=2 \int_{f_{\rm low}}^{f_{\rm high}}\frac{h_{,a} h^{*}_{,b}+h_{,b} h^{*}_{,a}}{S_n(f)} \,,
\ee
where $h_{,a} \equiv \partial h/\partial \lambda^a$ and $f_{\rm low, high}$ denote the integral's lower and upper frequency cutoffs. The one-$\sigma$ lower bounds on the $\lambda^a$ are given by $\sigma_a = \sqrt{\Sigma_{aa}}$, where the covariance matrix $\Sigma_{ab}=(\Gamma_{ab})^{-1}$ is the inverse of $\Gamma_{ab}$.

In computing $\Gamma_{ab}$ we choose a parameter set  
\be
\label{eq:lambdaA}
\lambda^a =  \left[t_{c},\, \phi_{c},\, \ln D_{L},\, \ln M_{c},\, \eta,\, \chi_{1z},\, \chi_{2z},\, \ln V_r \right]\,, 
\ee
where $(t_{c}, \phi_c)$ are the time and phase of coalescence, $D_{L}$ is the luminosity distance, $M_{c}$ is the source-frame chirp mass, $\eta$ is the symmetric mass ratio, and $\chi_{1z}$ ($\chi_{2z}$) is the primary's (secondary's) spin component along the orbital angular momentum. We ignore priors in $\Gamma_{ab}$; the $1\sigma$ errors are quite small for the high SNR cases we consider; priors do little to improve the precision. Masses are appropriately multiplied by $(1 + z)$ to account for the cosmological redshift $z$. We assume a flat universe with $H_{0}=67.90 {\rm (km/s)/Mpc}$, $\Omega_m=0.3065$, and $\Omega_{\Lambda}=0.6935$ \cite{Planck:2015fie}.  

We evaluate the Fisher matrix for a single 3G detector [the Cosmic Explorer (CE)~\cite{CEScience,CEScience19} or Einstein Telescope (ET)~\cite{ETScience11}], using noise PSDs in \cite{CEnET}. The {\tt IMRPhenomD} waveform is multiplied by a factor of $2/5$ to account for averaging over the inclination and angles entering the antenna pattern functions $F_{+,\times}$ \cite{Berti:2004bd,Arun:2006if,Damour:2000gg}. We also consider a LISA~\cite{Babak:2017tow} detector, ignoring its orbital motion and using the sky and polarization angle-averaged sensitivity in Ref.~\cite{RobsonLISAPSD}.\footnote{In the case of LISA, the {\tt IMRPhenomD} waveform amplitude needs to be multiplied by a factor $\sqrt{4/5}$ instead of $2/5$ to account for the inclination angle averaging \cite{RobsonLISAPSD}.} For CE (ET) we use $f_{\rm low} = 5\, {\rm Hz}\, (1 \, {\rm Hz})$ and $f_{\rm high} = 5000\, {\rm Hz}\, (10^4 \, {\rm Hz})$. For LISA $f_{\rm high} =0.1 {\rm Hz}$, and $f_{\rm low}$ is chosen according to Eq.~(2.15) of \cite{Berti:2004bd}, with a default cutoff $10^{-4}\, {\rm Hz}$ and 4 years of observation time.

\begin{figure*}[t!]
    \centering
    \includegraphics[width=1.0\textwidth]{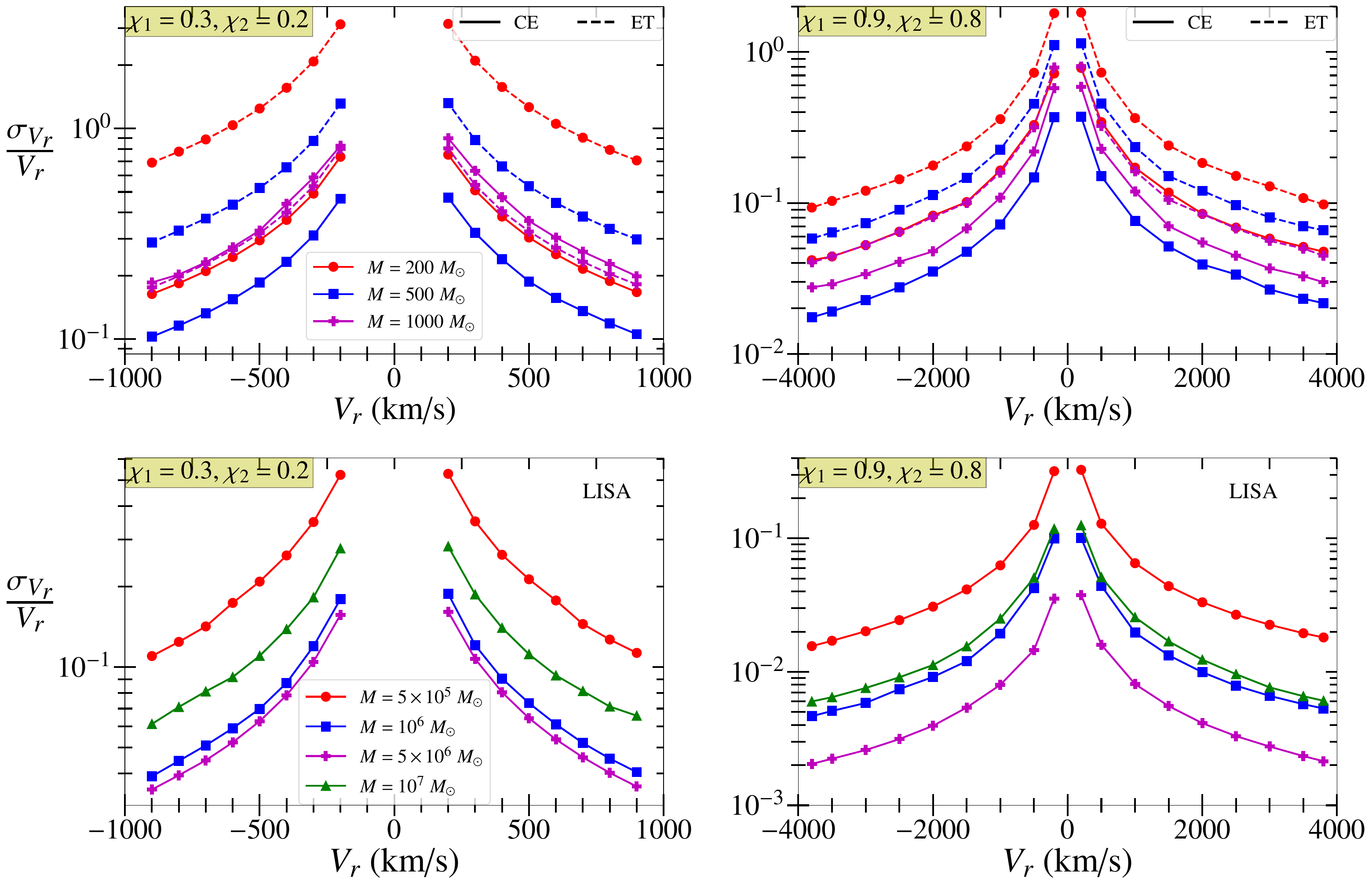}
    \caption{Fractional error in measuring the line-of-sight kick parameter $V_r$ as a function of $V_r$. The top two panels show measurements with the ground-based 3G detectors CE (solid curves) and ET (dashed); the bottom two panels show measurements with LISA. Black hole binaries have mass ratio $q=1.1$ with total (source-frame) masses indicated in the legend. Low-spin (left panels) and high-spin (right panels) configurations are considered, with the dimensionless spin parameters indicated. Sources are fixed at a luminosity distance of $1\, {\rm Gpc}$ ($z=0.198$) for ET/CE and $3\, {\rm Gpc}$ ($z=0.512$) for LISA. For the masses considered here, redshifted kicks ($V_r<0$) are better constrained than blueshifted ones because the ringdown signal is Doppler shifted to the more sensitive region of the noise curve; this accounts for the slight asymmetry of the curves on either side of $V_r=0$. SNRs range from $\sim 1500$--$4500$ (CE), $\sim 700$--$2200$ (ET), or $\sim 5000$--$22\,000$ (LISA).
    }
    \label{fig:kickerrors}
\end{figure*}
\section{Direct constraints on the LOS kick}
Figure \ref{fig:kickerrors} shows the fractional (1-$\sigma$) errors on the LOS kick parameter $V_r$ as a function of $V_r$. A range of total (source-frame) masses are shown in the intermediate mass BBH range (IMBBH, for CE/ET) and the supermassive BBH range (SMBBH, for LISA). We consider nearly equal-mass binaries ($q=1.1$) with low- and high-spin configurations (as in Fig.~\ref{fig:LOSkickCDF}). For a given $q$ and spin magnitudes, many spin orientations and LOS ($\hat{n}$) directions will yield the same $V_r$. In the low-spin configuration, the highest kicks are $V_{\rm kick} \sim 900\, {\rm km/s}$. When computing $\Gamma_{ab}$ we considered a specific spin orientation with $(\chi_{1z},\chi_{2z})=(0.06,0.03)$ that produced $V_{\rm kick} \sim 900\, {\rm km/s}$, varying the total masses and $V_r \in [-900, 900] \,{\rm km/s}$ to account for the LOS projection. In the high-spin case the max $V_{\rm kick} \sim 3800\, {\rm km/s}$, and we chose a configuration with $(\chi_{1z},\chi_{2z})=(0.51,0.44)$ and $V_r \in [-3800, 3800] \,{\rm km/s}$.

For low-spin binaries in the IMBBH mass range, kicks are typically $\lesssim 300\,{\rm km/s}$ and constraints from ground-based 3G detectors are moderate to poor ($\gtrsim 30\%$; cf.~Fig.~\ref{fig:kickerrors}, top-left panel). However, $\sim 20\%$ of high-spin, $q\approx 1.1$ binaries have kicks exceeding $\sim 1000\,{\rm km/s}$. In that case, 3G detectors like ET and CE can constrain $V_r$ to better than $20\%$ and in some cases to $\lesssim 2\%$. Note that CE typically provides better constraints than ET for most of the cases considered.

Since there are no observed IMBBHs with $M\gtrsim 200M_{\odot}$, the merger rates of these binaries are highly uncertain. To assess the prospects of constraining LOS kicks with 3G detectors, we need to estimate the number of expected IMBBH detections with sufficiently large kicks. We crudely estimate the detection rate of IMBBHs in the mass range $100M_{\odot}\hbox{--}1000M_{\odot}$ by summing the peak merger rates (middle panel of Fig.~1 in \cite{Fragione:2022avp}) and multiplying by the comoving volume out to $z\sim 0.2$ (where ringdown signals will be well measured). This yields $\sim (29 \,{\rm Gpc}^{-3} \,{\rm yr}^{-1}) (2.5\,{\rm Gpc}^{3}) \sim 73$ IMBBHs per year observed by 3G detectors. In the absence of constraints on the mass and spin distributions of the IMBBH population, we sample binaries with random spin orientations and observer locations ($\hat{n}$), uniformly choosing mass ratios and spins in the range $q\in[1,20]$ and $\chi_{1,2} \in [0,0.99]$. After applying this sample to the recoil formula~\cite{Lousto:2012su,Lousto:2012gt}, we find that $\sim 6\%$ of these IMBBH binaries will have  $|V_r|\geq 1000 \, {\rm km/s}$. Since these binaries are nearly equal mass with high spins, they will have $V_r$ precisely measured (Fig.~\ref{fig:kickerrors}, top right). Hence, assuming this merger rate model is representative of the true IMBBH population, we expect to observe $\sim 6\% \times 73 \sim 4$ IMBBH per year for which $V_r$ is well constrained. 

LISA provides better constraints on the LOS kicks of SMBBHs (due to the larger ringdown SNR compared to 3G ground-based detectors). In the low-spin case, nearly equal-mass binaries constrain $V_r$ to $\lesssim 20\%$ for $M\gtrsim 10^6 M_{\odot}$ (cf.~bottom left of Fig.~\ref{fig:kickerrors}). In the high-spin case $V_r$ is constrained to $\lesssim 6\%$ in most cases for all considered masses, and to better than $1\%$ in some cases. Note that higher spin binaries are better constrained due to both their higher kicks and higher SNRs.

We also consider LOS kick constraints on observed systems. GW200129$\_$065458 \cite{LIGOScientific:2021djp} shows evidence for spin-orbit precession~\cite{Hannam:2021pit} and has an inferred kick speed of $1542^{+747}_{-1098}$ km/s \cite{Varma:2022pld}. Assuming a LOS kick of $1000\, {\rm km/s}$, CE (ET) could constrain $V_r$ from a system with consistent parameters [$q=2,\,M=62 M_{\odot},\, z=0.2,\, (\chi_{1z},\chi_{2z})=(0.9,0.5)$ \cite{Varma:2022pld,Hannam:2021pit}] to within $\sim 30\%$ ($\sim 70\%$).  Quasars with spectra Doppler-shifted relative to their host galaxies are also candidates for kicked SMBHs \cite{Komossa:2008qd, Shields2009ApJ, Boroson2009Natur, Civano2010ApJ, Robinson2010ApJ}. For SDSS J105041.35+345631.3 \cite{Shields2009ApJ} [with parameters $V_{r} \approx +3500\, {\rm km/s}$, $z = 0.272$, $M=10^{7.5}\, M_{\odot}$, $q=1.1$, and $(\chi_{1z},\chi_{2z})=(0.51,0.44)$], LISA can measure the observed LOS kick with $\approx 1\%$ precision. 

\begin{figure}[th]
\centering
\includegraphics[scale=0.46]{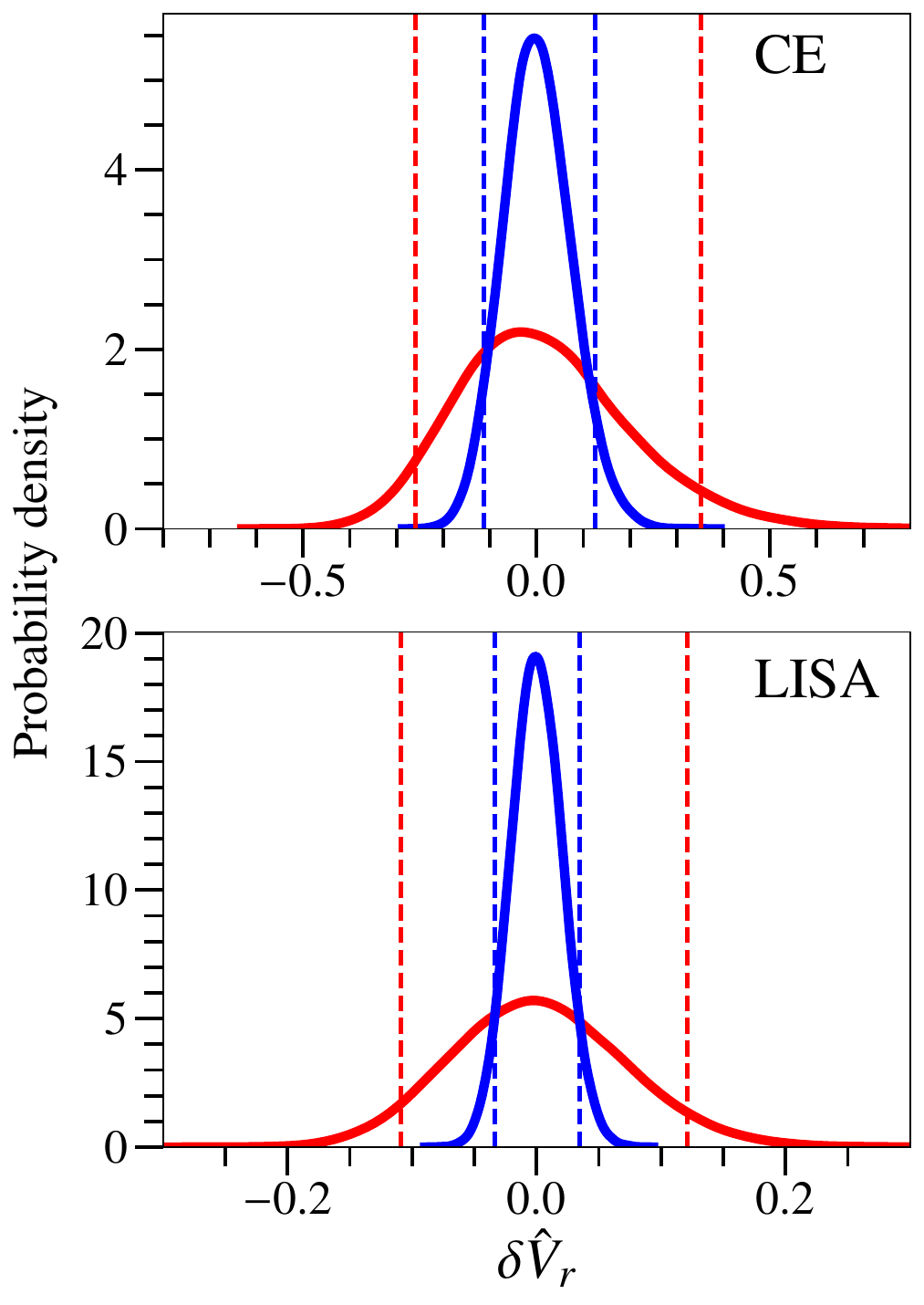}
\caption{\label{fig:kicktgrbound}Bounds on the dimensionless null test parameter $\delta \hat{V}_r$ as set by CE (top) or LISA (bottom). The parameter $\delta \hat{V}_r$ measures the consistency between the line-of-sight kick directly measured via the ringdown signal and that inferred via the system parameters and NR simulations of the kick. Dashed vertical lines are 90\% confidence intervals. Results show the same high-spin (blue) and low-spin (red) configurations as previous plots, with $q=1.1$. The high-spin case uses spin vectors (defined at a reference time of $-100 M$) of $\vec{\chi}_1 =(-0.74, 0.04, 0.51)$, $\vec{\chi}_2 =(0.67, 0.02, 0.44)$, with $\iota=4\pi/7$ and a resulting LOS kick $V_r=1004\, {\rm km/s}$. For the low-spin case: $\vec{\chi}_1 =(0.14, -0.26, 0.06)$, $\vec{\chi}_2 =(-0.08, 0.18, 0.03)$, and $\iota=9\pi/13$, resulting in $V_r=-506\, {\rm km/s}$. In the CE case $M=500 M_{\odot}$, $D_L= 1\,{\rm Gpc}$, and the low (high) spin SNR is $2800$ ($1700$); for LISA, $M=10^6 M_{\odot}$, $D_L= 3\,{\rm Gpc}$, with low (high) spin SNR of $5800$ ($2900$).
}
\end{figure}

\section{\label{sec:GRtest}Testing GR via linear momentum balance}
The precise measurement of LOS kicks (discussed above) can be leveraged to formulate a null test of GR based on the consistency between the inferred and directly measured LOS kick. This compares the directly measured LOS kick $V_r$ (from the Doppler-shifted ringdown as above) with the LOS kick inferred via the estimated system parameters combined with NR models for the kick ($V_r^{\rm NR}$).\footnote{Here we assume that the system parameters are inferred via the entire BBH signal (inspiral, merger, ringdown/IMR); this improves the constraints. A variant of this test could compare NR fit predictions for the kick using pre- and post-merger parts of the waveform. This is similar to the IMR consistency test~\cite{Ghosh2016,Hughes:2004vw}. 
} 

To formulate this null test we define two new variables $(\bar{V}_r, \delta \hat{V}_r)$ and their inverses:
\begin{align}
\bar{V}_r &\equiv \frac{1}{2} (V_r + V_r^{\rm NR}) \,, \;\;\;\;
\delta \hat{V}_r \equiv \frac{V_r-V_r^{\rm NR}}{\bar{V}_r}\,,\\
{V}_r &= \bar{V}_r  (1+ \frac{1}{2} \delta \hat{V}_r) \,, \;\;\;\; 
{V}_r^{\rm NR} = \bar{V}_r  (1- \frac{1}{2} \delta \hat{V}_r) \,.\label{eq:dVdef}
\end{align}
Given observational data $d$, our goal is to compute the posterior probability distribution $p(\delta \hat{V}_r|d)$ on the null variable $\delta \hat{V}_r$ and test its consistency with zero. To compute this we marginalize the joint probability distribution function (PDF) $p(\delta \hat{V}_r, \bar{V}_r|d)$ for observing $\delta \hat{V}_r$ and $\bar{V}_r$. This is related to the joint PDF $p(V_r|d) p(V_r^{\rm NR}|d)$ for measuring $V_r$ and $V_r^{\rm NR}$ via
\be
\label{eq:dVprob}
p(\delta \hat{V}_r|d) = \int p(V_r|d) \, p(V_r^{\rm NR}|d) \, \bar{V}_r \,d\bar{V}_r \,,
\ee
where the factor $\bar{V}_r$ arises from the Jacobian of the coordinate transformation $\partial (V_r, V_r^{\rm NR})/\partial (\delta \hat{V}_r, \bar{V}_r)$.

To compute $p(V_r^{\rm NR}|d)$ we use a parameter set $\lambda^i=\{ t_{c},  \phi_{c}, \ln \, D_{L}, \ln \, M_{c}, \eta, \chi_{1x}, \chi_{1y}, \chi_{1z}, \chi_{2x}, \chi_{2y}, \chi_{2z},$ $\cos \iota, \Phi\}$, where $(\iota, \Phi)$ are the angles $\hat{n}$ makes relative to the binary's orbital angular momentum at a reference time. (Note that $\iota \equiv \Theta$.) Using the {\tt IMRPhenomXPHM} waveform \cite{IMRPhenomXPHM} (which accounts for higher modes and precession) and the resulting Fisher matrix $\Gamma_{ij}$, we construct a Gaussian PDF  $p(\lambda^i) \propto e^{-\frac{1}{2} \Gamma_{ij}(\lambda^i-\lambda^i_{\rm inj})(\lambda^j-\lambda^j_{\rm inj})}$, where $\lambda^i_{\rm inj}$ are the injected parameter values. (No priors are applied.) We integrate over the entire frequency range encompassing the inspiral, merger, and ringdown as discussed earlier; for massive sources with significant merger/ringdown SNRs, this improves our overall parameter constraints (relative to integrating only over the inspiral signal). The right ascension, declination $(\alpha, \delta)$, and polarization angle $\psi$ of the source are kept fixed. For simplicity in this initial investigation, we choose $\alpha=\delta=\psi=\Phi=\pi/3$ in all cases. We feed samples from this multidimensional Gaussian into $V_r^{\rm NR} = \vec{V}_{\rm kick}^{\rm NR} \cdot \hat{n}$, where $\vec{V}_{\rm kick}^{\rm NR}$ is evaluated via NR surrogate models \cite{Varma:2018aht,Varma:2019csw}. This produces a normalized PDF for $p(V_r^{\rm NR}|d)$. Note that {\tt IMRPhenomXPHM} does not build in a Doppler shift; rather, we use that waveform to \emph{predict} the Doppler shift via $V_r^{\rm NR}$.

We then construct $p(V_r|d)$ by marginalizing the PDF $p(\lambda^a)$ computed from $\Gamma_{ab}$ with parameters $\lambda^a$ [Eq.~\eqref{eq:lambdaA} and Fig.~\ref{fig:kickerrors}]. This yields the 1D PDF $p(V_r|d) \propto e^{-\frac{1}{2} (V_r - V_r^{\rm inj})^2/\sigma_{V_r}^2}$.  To evaluate the integral in Eq.~\eqref{eq:dVprob} we construct a discrete grid of values for $\delta \hat{V}_r$ and $\bar{V}_r$, compute the corresponding $(V_r, V_r^{\rm NR})$ via \eqref{eq:dVdef}, and discretely sum the integrand over the relevant points. 

The above procedure results in the PDF $p(\delta \hat{V}_r|d)$ shown in Fig.~\ref{fig:kicktgrbound}. This method is applied to a near-equal mass ($q=1.1$) binary observed with a single CE detector or LISA. (We ignore LISA's orbital motion and angle average over $\alpha$, $\delta$, and $\psi$ in that case.) At $90\%$ confidence our low-spin (moderate LOS kick) binary constrains the null variable to $\delta \hat{V}_r = 0^{+0.35}_{-0.26}$ ($0^{+0.12}_{-0.11}$)  with CE (LISA). For our high-spin (high LOS kick) binary, the constraint improves to $\delta \hat{V}_r = 0^{+0.13}_{-0.11}$ ($0^{+0.03}_{-0.03}$)  with CE (LISA).

\section{Conclusions}
We have demonstrated the direct measurability of BBH kicks via their Doppler-shifted ringdown signal. Depending on the binary spin (and kick magnitude), the line-of-sight kick component can be constrained to $\sim 2\%$ to $\sim 30\%$ with 3G ground-based detectors, and to $\lesssim 1\%$ precision in some cases with LISA. 

We also proposed a new test of GR based on the consistency between the inferred and directly measured line-of-sight kick. A null variable that quantifies this consistency can be measured with $\sim 10\%$ to $\sim 30\%$ precision with CE and $\sim 3\%$ to $\sim 12\%$ precision with LISA. This impressive precision suggests that the proposed \emph{kick consistency test} could be an important science target for LISA and 3G detectors. Just as the Hulse-Taylor binary first showed that GWs carry energy and angular momentum, these measurements could similarly provide direct evidence that GWs carry linear momentum as GR predicts.

Our goal has been to provide a proof-of-concept for this consistency test. A simulated injection campaign using more sophisticated waveform models, a wider range of binary parameters, and an examination of systematic errors, will better quantify the utility of this proposed GR test. In particular, future work will need to more carefully understand how the kick---along with the final mass and spin--can be extracted from a multimode ringdown signal.


\acknowledgements
K.~G.~A.~acknowledges support from the Department of Science and Technology and Science and Engineering Research Board (SERB) of India via the following grants: Swarnajayanti Fellowship Grant No. DST/SJF/PSA-01/2017-18 and MATRICS Grant (Mathematical Research Impact Centric Support) No. MTR/2020/000177. K.~G.~A. thanks Montclair State University for hospitality during the early stage of this work. K.~G.~A and P.~M. acknowledge the support of the Core Research Grant No. CRG/2021/004565 of the Science and Engineering Research Board of India and a grant from the Infosys foundation. M.~F.~acknowledges NSF support via CAREER Award No. PHY-1653374. We thank Keefe Mitman for a helpful discussion that clarified issues related to NR simulations and kicks. We also thank the referees for helpful comments that improved the presentation.
\bibliography{ref-list}


\end{document}